\documentclass{article}
\usepackage[utf8]{inputenc}
\usepackage{authblk}
\usepackage{setspace}
\usepackage[margin=1.25in]{geometry}
\usepackage{graphicx}
\graphicspath{ {./figures/} }
\usepackage{subcaption}
\usepackage{amsmath}

\usepackage[style=nejm, 
citestyle=numeric-comp,
sorting=none]{biblatex}
\title{Fully Integrated Vacuum-based Quantum Random Number Generator}

\author[1$\dag$]{Xin Hua }
\author[2$\dag$]{Yiming Bian}
\author[1]{Ying Zhu}
\author[2]{Jiayi Dou}
\author[2,3]{Jie Yang}
\author[1]{Shengxiang Zhang}
\author[1]{Jie Yan}
\author[1]{Min Liu}
\author[1]{Daigao Chen}
\author[2]{Song Yu}
\author[3]{Bingjie Xu}
\author[2*]{Yichen Zhang}
\author[1*]{Xi Xiao}

\affil[1]{National Information Optoelectronics Innovation Center (NOEIC), China Information and Communication Technologies Group Corporation (CICT), Wuhan, 430074, China}
\affil[2]{State Key Laboratory of Information Photonics and Optical Communications, Beijing University of Posts and Telecommunications, Beijing, 100876, China}
\affil[3]{ Science and Technology on Communication Security Laboratory, Institute of Southwestern Communication, Chengdu, 610041, China}
\affil[*]{Address correspondence to: zhangyc@bupt.edu.cn, xiaoxi@noeic.com}
\affil[$\dag$]{These authors contributed equally to this work.}

\date{}

\begin{document}

\maketitle

\begin{abstract}
Quantum random number generators play a crucial role in securing high-demand information contexts by producing true random numbers. Nevertheless, the large volume and high-cost limit their widespread use. Here, we propose a system on chip that fully leverages the advantages of different photonic integrated platforms, where the interference optical paths and photodiodes are integrated on a standard silicon process, while the laser source on-chip is realized on a III-V platform. Using micro-lens coupling package technology, which contributes to a topnotch coupling loss lower than 2\,dB, the components on different platforms are combined and packaged with the amplifier circuits in a 42\,mm $\times$ 24\,mm footprint in a butterfly form. This complete miniaturized and cost-effective entropy source enables outputting a vacuum noise signal with a 3\,dB bandwidth of over 500\,MHz. After sampling and post-processing, a random number generation rate of up to 6.57\,Gbps is achieved. The results show a feasible way of overcoming the laser integration problem with silicon-based integrated quantum photonics. Foreseeable, commercial applications on a large scale are significantly promoted.
\end{abstract}


\section{Introduction}
Random numbers are widely used in various applications, such as gambling, simulations \cite{ferrenberg1992monte}, and cryptography \cite{gennaro2006randomness}. Although computer algorithms or classical noisy systems \cite{uchida2008fast} can easily produce random numbers, they all lack provable true randomness, since the nature of classical systems is deterministic. For certain applications requiring true random numbers, such as quantum key distribution \cite{PTPQKDRMV2020, AdvInQC, zhang2024continuous}, the classical methods are not suitable and may introduce security loopholes \cite{bouda2012weak}. In these scenarios, ensuring true randomness is rather important. 

Thanks to the intrinsic randomness in quantum mechanics, a quantum system is effective for generating strings cannot be predicted, namely quantum random number generator (QRNG) \cite{Ma16, herrero2017quantum}. Today, there are numerous entropy sources \cite{Zhou2019quantum, Huang2021}, within them, the principles based on photonics, including photon number statistics \cite{Ren2011}, amplified spontaneous emission (ASE) \cite{Williams10, Wei12, liu2013implementation, martin2015quantum}, phase noise \cite{qi2010high, guo2010truly, xu2012ultrafast, abellan2014ultra, nie2015generation, liu2016117}, and the vacuum fluctuation \cite{gabriel2010generator, shen2010practical, symul2011real, haw2015maximization, xu2019high, zheng20196}, have drawn researchers' attention since the scheme of entropy source is easily built using the off-the-shelf bulk optics with remarkably high generation rate.

For the wide use of QRNG, several obstacles should be addressed: cost, package size and form. Photonic integrated circuits (PIC) are a promising way to overcome all limitations \cite{wang2020integrated}. To date, the process technique for different PIC platforms, i.e., silicon-on-insulator (SOI), lithium niobate, \uppercase\expandafter{\romannumeral3}-\uppercase\expandafter{\romannumeral5} compounds, are becoming more and more mature, one can efficiently manipulate, and detect photonic quantum state \cite{wang2020integrated, laurent2023PRXquantum}. This pushes optoelectronic applications, including QRNG, to a photonic fashion. Through schemes of ASE and phase noise, one can achieve high generation rates; however, implementations based on PIC are challenging. For instance, the ASE scheme normally needs the amplifier module of laser power, while the phase noise needs a stabilized unbalanced Mach-Zehnder interferometer. Both of these make the structure of the source complicated, potentially increasing the device cost. In contrast, the vacuum fluctuation-based entropy takes advantage of the commercial integrated coherent receiver widely used in the classical telecommunication domain \cite{Hu2021, Chen2022}, which includes the balanced homodyne detector (BHD), which is advantageous in opening the avenue to commercialization. 
 
The balanced homodyne detection module can be fabricated on \uppercase\expandafter{\romannumeral3}-\uppercase\expandafter{\romannumeral5} or SOI platforms. The \uppercase\expandafter{\romannumeral3}-\uppercase\expandafter{\romannumeral5} material, such as indium phosphide (InP), allows to integrate light source, leading to a monolithic local oscillator-included homodyne detection\cite{abellan2016quantum, roger2019real, chrysostomidis2023long}. However,  \uppercase\expandafter{\romannumeral3}-\uppercase\expandafter{\romannumeral5} platform is currently cost-irreducible due to the low fabrication yield, while the SOI platform is outstanding with mature fabrication technique. The compatibility with the CMOS processes makes it a promising candidate for large-scale integrated applications, potentially supporting the QRNG source adoption of more scenarios, such as the silicon-based QKD encoder \cite{zhang2019integrated, Wei:23}, of which the randomness of quantum state encoding will be guaranteed. Considering the low-cost and high-density integration advantages,  silicon-based QRNG is appealing in terms of commercial application and will likely be widely deployed in an information-secure network.

Though SOI enables the realization of the BHD module, it is impractical for laser fabrication due to the indirect bandgap of silicon material. The works dedicated to this usually rely on a bulk laser device to offer local oscillator signal \cite{raffaelli2018homodyne, Bai2021, PRXQuantum.4.010330}. This unquestionably hampers the miniaturization of QRNG. In addition to the optical circuit, some peripheral electrical circuits need to be implemented to convert the differential photocurrent signal into a measurable voltage signal, which should be effectively captured by an analog-to-digital converter (ADC). The electrical circuits typically comprise amplifiers and other auxiliary components, such as filtering circuits \cite{Bai2021}. It is crucial to minimize the size of the electrical circuit to further reduce the overall size of QRNG device. Furthermore, the electrical circuits should solely amplify the quantum shot noise without introducing additional noise, being part of classical noise, which would degrade the randomness performance of QRNG.

In this work, we designed and fabricated BHD based on a standard silicon process. Based on a heterogeneous integration technique, we packaged indium phosphide distributed feedback (DFB) laser chip, silicon-based BHD, and low-noise electrical amplifiers in a butterfly form with a small footprint of  42 mm$\times$24 mm. Using a 12-bit-ADC oscilloscope with sampling rate of 625\,MSa/s, a random number generation rate of 6.57\,Gbps can be achieved. 
Remarkably, our QRNG device was achieved with off-the-shelf commercial components, including the laser chip and electrical amplifiers, ensuring a miniaturized highly reliable cost-effective electrical in and out form QRNG. 

\section{Methods}\label{sec2}

\begin{figure}[h]
    \centering
    \includegraphics[width=\textwidth]{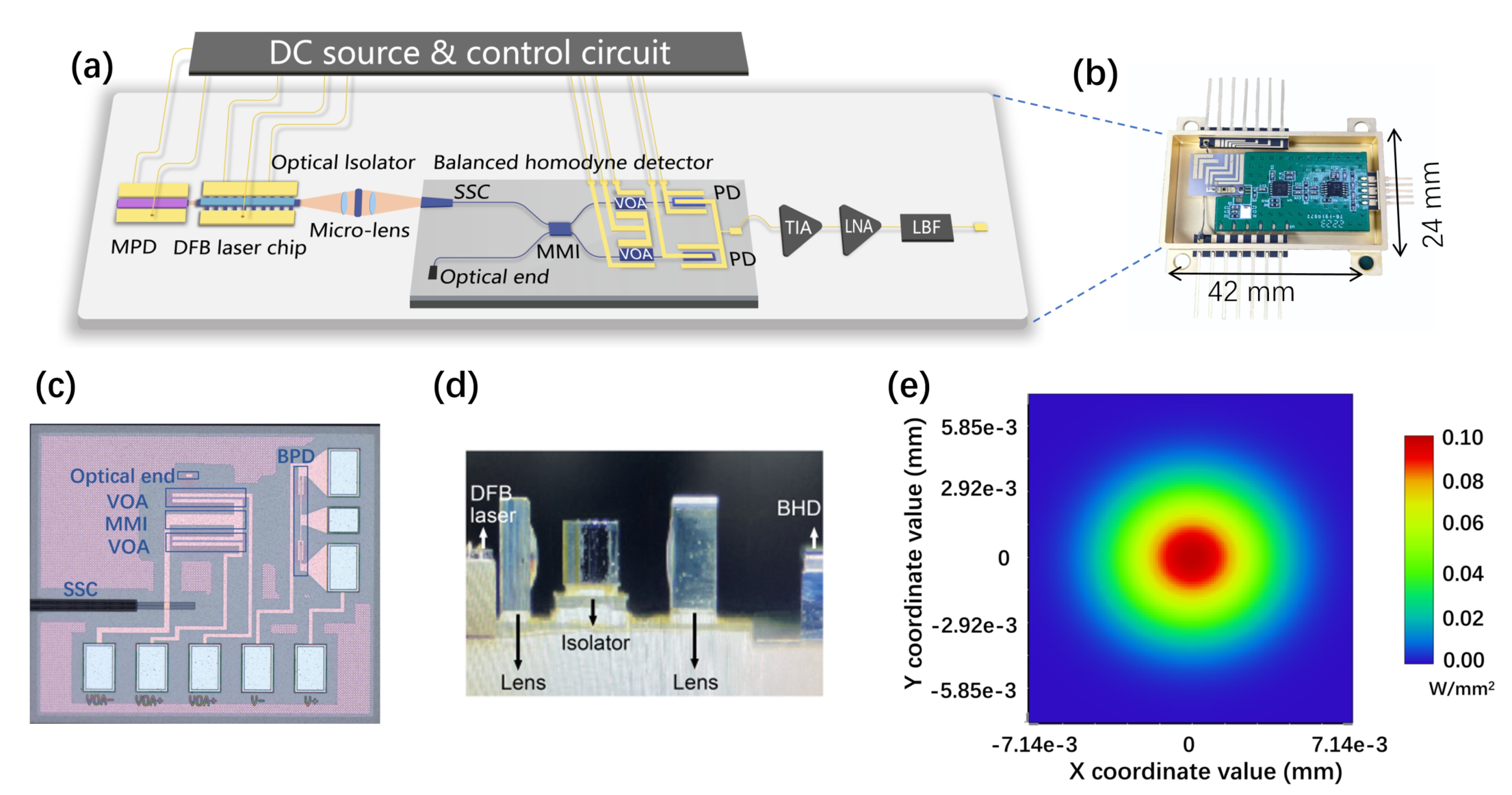}
    \caption{(a) The inside schematic diagram and (b) photograph of QRNG entropy. (c) The microscope image of the BHD chip. (d) The microscope image of the coupling structure between the DFB laser and BHD. (e) The simulation result of the beam field distribution in free space prior to entering the BHD chip. MPD: monitor photodetector, SSC: spot-size converter, MMI: multimode interferometer coupler, VOA: variable optical attenuator, PD: photodetector, TIA: transimpedance amplifier, LNA: low noise amplifier, LBF: low bandpass filter. } 
    \label{fig:QRNG_entropy}
\end{figure}

The chip-based QRNG depicted in Fig.\ref{fig:QRNG_entropy}a comprises a DFB laser diode, a germanium-on-silicon (Ge-Si) BHD (see Fig.\ref{fig:QRNG_entropy}c), a TIA, a LNA, and a LBF utilizing heterogeneous integration technique. The DFB laser is realized by implementing a commercial InP chip, of which the optical output power is tunable by applying different operation currents.  The typical output power can reach 19.5\,dBm when operating with 200\,mA at 25$^{\circ}$C. The corresponding linewidth of the laser is 350\,kHz. Notably, to ensure that the entropy source works properly and continuously, two feedback circuits respectively focus on the temperature and operating current of the laser, which are externally implemented to control the source in real time. A negative temperature coefficient thermistor is attached to the laser chip and a ceramic heatsink is used to mount the laser chip, the thermoelectric cooler (TEC) module stabilizes the temperature of laser chip based on the feedback temperature value. A commercial MPD is placed behind the DFB laser chip, connected to the operating current control circuit. By monitoring the power of backward transmitted laser beam from the laser cavity, the operating current is tuned to maintain a constant output power of the laser.

Thanks to Ge-Si PDs \cite{Hu2021}, a BHD can be realized on a monolithic standard SOI wafer with a 220\,nm silicon layer, and a 3\,$\mu$m buried silica oxide layer.
The width of single mode silicon waveguide is 450\,nm. $2\times2$ MMI is designed ensuring a balanced splitting ratio. VOA is based on the forward carrier injection PIN junction, offering an attenuation ranging from 0 to 40\,dB. The balanced photodetector consists of two co-directional Ge-Si PDs, connected in series. With a reverse bias applied on the BPD, the common electrode outputs the differential photocurrent RF signal. In our case, the two PDs reveal near the same electro-optical characteristics, such that the dark current is lower than 50\,nA, and the typical responsivity is 0.85\,A/W  at 1550\,nm when balanced photodetector is biased with -2\,V. The full footprint of this SOI chip is 700\,$\mu$m$\times$1200 $\mu$m. 

The laser beam is coupled into the BHD module using a telescope alignment architecture that incorporates two micro-lenses (see Fig.\ref{fig:QRNG_entropy}d). The primary function of the first miro-lens, following the laser chip, is to facilitate beam collimation, whereas the second lens is responsible for beam convergence. The principle of coupling between InP laser and silicon BHD chips is to change the mode field size and numerical aperture of laser beam relying on the group of micro-lenses. Using the Zemax simulation tool, we optimized the coupling ratio by fine-tuning structural parameters, including the curvature and diameter of the micro-lenses, as well as the distance between them. Following mode size conversion, the light field distribution in free space prior to the SSC is illustrated in Fig.\ref{fig:QRNG_entropy}e. The beam widths in both orthogonal directions, measured at $\frac{1}{e^2}$ of the peak intensity, are approximately 4.5 µm, which is smaller than the 9 µm mode size of the SSC. This ensures that the majority of the beam power enters the SOI chip, achieving a state-of-the-art coupling loss of 2 dB from the laser output to the SSC coupler edge. In accordance with the GR-468 standard, the reliability of this packaging was rigorously tested, including 1000 hours of storage at a high temperature of 85°C and 72 hours of storage at a low temperature of -40°C. Additionally, the structure underwent 500 cycles of high-low temperature cycling, with the coupling efficiency fluctuation remaining below 1$\%$. To mitigate the impact of back-reflected light on laser stability, a C-band optical isolator was integrated between the two micro-lenses.

The light playing a role of local oscillator, is coupled into chip via SSC, then equally distributed into two parts by a balanced MMI. Another input of MMI connects to optical end ensuring a vacuum state. Two VOAs are respectively employed on two outputs of MMI, to compensate the unbalance introduced due to the fabrication imperfection. Two PDs in series detect lights passing through VOAs. The common electrode of two PDs outputs a differential photocurrent RF signal via wire-bonding technique, which is then converted into a voltage signal by a commercially available low-noise TIA with a wide analog bandwidth of around 600\,MHz, a gain of around 80\,dB$\Omega$. To ensure accurate and effective data collection by the ADC, a commercial wideband LNA with a gain of 34 dB is employed. The two cascade amplifiers can output an electrical power level higher than -40 dBm, which falls within the measurement range of the oscilloscope (Keysight MSOS404A) used in the experiment. To suppress noise out of range of the bandwidth of TIA, the signal is filtered by a LBF with a bandwidth of 600\,MHz. In Fig.\ref{fig:QRNG_entropy}a, all the optical components are glued on a ceramic facing with a tiny size of 2\,mm$\times$5\,mm, while the whole components excluded direct current sources and control circuits, are butterfly packaged within a footprint of 42\,mm$\times$24\,mm. 

To make the QRNG work, we designed a controller board to apply the corresponding voltage signal to different pins of the butterfly module. Typically, the MPD was applied a reverse bias of 5\,V, and gave a readout of photon-current, which helped to tune the applied voltage in real-time maintaining constant the output laser power via a feedback circuit connecting to the laser diode.
The voltage applied to TEC was also tunable based on a temperature-feedback circuit for stabilizing the working temperature of the laser chip. The anode and cathode of BPD were respectively applied 1\,V, -1\,V. For electrical amplifiers, including TIA and LNA, the applied voltages are, respectively, 3.3\,V and 5\,V. All the voltages are supplied by the controller board powered by a single external power source of 12\,V. 

\section{Results and Discussion}
\subsection{Noise Power Spectrum Density Measurement}
\begin{figure}[h]
    \centering
    \includegraphics[width=\textwidth]{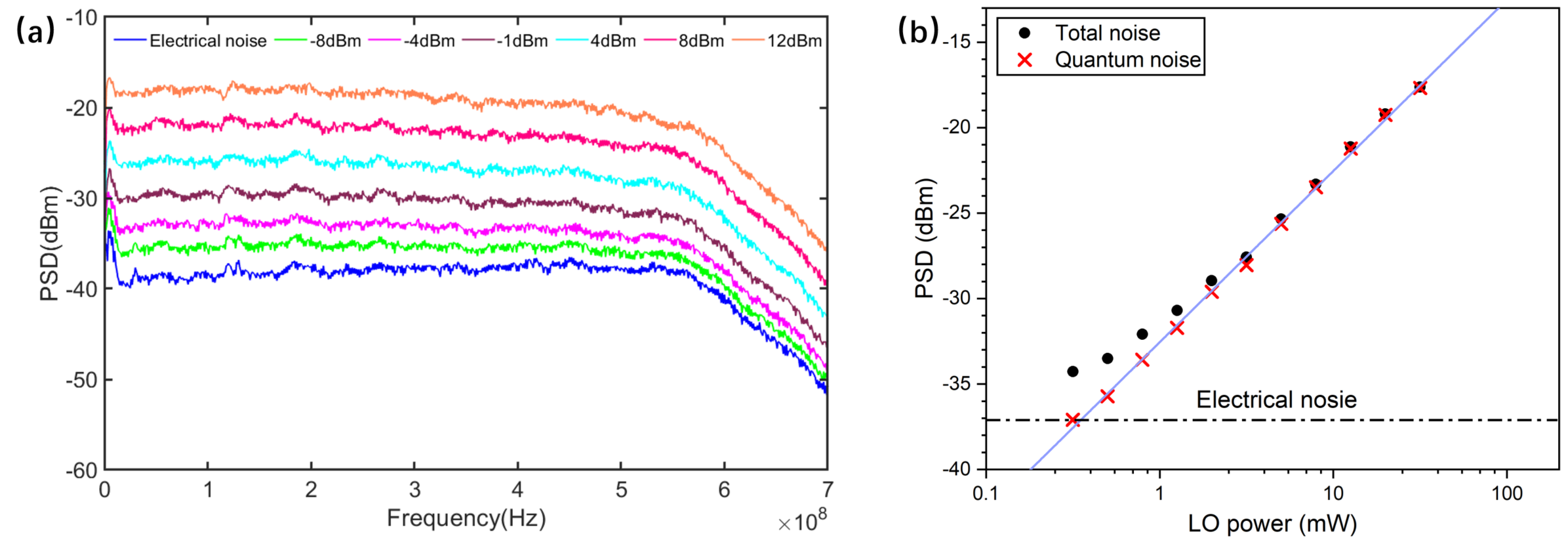}
    \caption{(a) Average PSD at the output of LNA when the laser power is tunned. (b) Average PSD (black dots) measured at 200\,MHz for different incident LO powers. The black dot line represents the electrical noise, and the blue line represents the PSD fit for quantum shot noise (red crosses). }\label{fig:PSD}
  \end{figure}

Since the vacuum fluctuation measurement is based on the BHD, whose performance directly determines the specification of QRNG, we separately characterized the performance of BHD. Firstly, the PSD of vacuum fluctuation for different incident laser powers was measured. A 1550\,nm continuous laser (Santec TSL550) was incident into the BHD chip through a single-mode fiber coupled to the LO port. Between them, a polarization controller was implemented to choose the TE mode for maximizing the power passing through the silicon chip. The output of the QRNG entropy source was measured by a signal analyzer (Keysight N9030B). Varying the incident laser power, different PSDs, as shown in Fig. \ref{fig:PSD}a were obtained. From the results, the PSD increases with the LO pump power, and the PSD keeps nearly flat with a 3\,dB bandwidth of over 500\,MHz, approaching the TIA bandwidth. Note that since the maximum output power of laser device is 15\,dBm, the PSD test was not further performed until the PDs went to saturation. The results in Fig. \ref{fig:PSD}a take into account of the 3\,dB loss from the output of laser to the SSC edge of the BHD chip.
 
The noise PSD consists of quantum shot noise and electronic noise. The PD intrinsic electronic noise and the amplification circuit mainly contribute to the electrical noise. To estimate the quantum noise of QRNG, we extracted the average values of PSD at 200\,MHz for different incident LO powers in Fig. \ref{fig:PSD}a, the PSD of electrical noise was obtained when the laser was turned off. According to the PSDs of total noise and electrical noise, the PSD of quantum shot noise was conducted, of which the fit was linear in function of LO power (see Fig.\ref{fig:PSD}b), being compatible with the theoretical prediction for vacuum state fluctuation based QRNG scheme. When the LO power was set to 15\,dBm, the quantum-to-classical noise clearance was around 20\,dB before the PDs reached saturation.

\subsection{Random Number Estimation}

\begin{figure}[h]
    \centering
    \includegraphics[width=\textwidth]{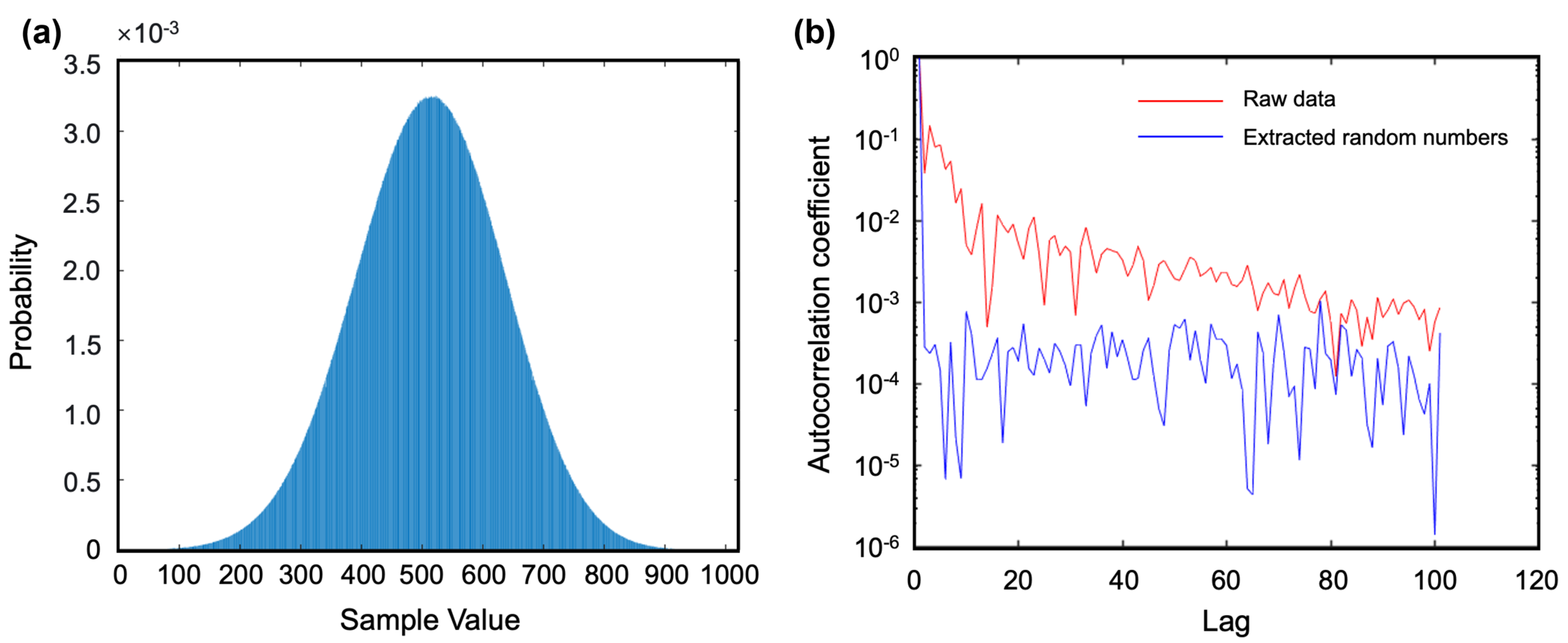}
    \caption{(a) The histogram of the sampled raw data. (b) The autocorrelation coefficient of raw random data (red) and processed data (blue).}\label{fig:histo_data}
\end{figure}

To extract random numbers,the noise signal from the entropy source was firstly sent to an oscilloscope (Keysight MSOS404A) for sampling, then post-processed by a computer for randomness distillation. The randomness is characterized by the min-entropy of the quantum noise.
The electronic noise data is achieved with LO off, and the total noise data, also called the raw data, is achieved with LO on. In our case, we set the laser diode to work with a medium current of 120\,mA, which outputs a corresponding power of around 14\,dBm. Considering the loss from the laser diode to the BHD, the LO power is 12\,dBm. The oscilloscope was set to a sampling rate of 625\,MSa/s with a 12-bit ADC converter. The measured noise data was normalized and shown in Fig.\ref{fig:histo_data}a satisfying a Gaussian distribution. 

The min-entropy determines the random information amount that can be extracted from the noise. The calculation of min-entropy considers the worst-case where the adversary Eve can fully control the classical noise. Denote the measured signal as $M$, and Eve’s knowledge of the classical noise as $E$, the min-entropy can be conducted by
\begin{equation}
    H_{min}(M|E)=-log_2[erf(\frac{\delta}{2\sqrt{2}\sigma_Q})]
\end{equation}
Here, the modeled classical noise includes the electronic noise of the detector and the sampling noise of the ADC card. Specifically, we assume that the shot noise and the classical noise are independent, thereby the standard deviation of the quantum fluctuation, $\sigma_Q$, can be conducted by $\sigma_M^2=\sigma_Q^2+\sigma_C^2$, where $\sigma_M^2$ and $\sigma_C^2$ are respectively the variance of the sampled data, and of the classical noise. $\delta$ represents the sampling precision, which can be obtained by $\delta=\frac{2R}{2^n}$, where $2R$ is the input voltage range of the ADC card, and $n$ is the sampling bits. Denote the standard deviation of the calibrated electronic noise as $\sigma_E$, $\sigma_C^2$ can be written as $\sigma_C^2=\sigma_E^2+2(\delta/n)^2$, where $(\delta/n)^2$ represents the variance of quantization errors that exist in the discretization of $M$ and $E$. In experiments, $R=1.5\,V$ and $n=12$. $\sigma_E^2$ can be calibrated from the electronic noise of the homodyne detector without optical signal input, and $\sigma_M^2$ is achieved from the measurement data when LO is on. The calculated $\sigma_Q$ is around 0.426 V, and the min-entropy obtained is 10.51 bits. After the Toeplitz hashing operation, the final true random number data with a generation rate of 6.57 Gbps is achieved.

The randomness before and after the hashing operation is revealed by the autocorrelation coefficient, as shown in Fig.\ref{fig:histo_data}b. The weak autocorrelation of the raw data is effectively suppressed after post-processing.  
The final random numbers pass all the test items of a standard NIST statistical test.

\begin{table}[h!]
\caption{Comparison of State-of-the-Art Chip-based QRNG}
\label{tab:state-of-the-art}
\begin{tabular}{c c c c c c}
\hline
Ref        & Type         & Laser source  & Optical path  & Detection        & Speed (Gbps) \\ \hline

\cite{Bai2021}  & Vacuum state & Bulk          & On-chip (SOI) & On-chip (InGaAs) & 18.8         \\
\cite{PRXQuantum.4.010330}   & Vacuum state & Bulk          & On-chip (SOI) & On-chip (SOI)    & 100          \\
\cite{raffaelli2018generation} & Phase noise  & Bulk          & On-chip (SOI) & On-chip (SOI)    & 2.8          \\
\cite{chrysostomidis2023long}  & Phase noise  & \textbf{On-chip (InP)} & On-chip (InP) & On-chip (InP)    & 6.11         \\
Our work   & Vacuum state & \textbf{On-chip (InP)} & On-chip (SOI) & On-chip (SOI)    & \textbf{6.57}    \\ \hline

\end{tabular}
\end{table}

Table \ref{tab:state-of-the-art} shows comparisons of state-of-the-art chip-based QRNG in literature. It is evident that the majority of works, such as \cite{Bai2021, PRXQuantum.4.010330, raffaelli2018generation} are prevented by laser integration. \cite{chrysostomidis2023long} is the first and only work that achieves monolithic chip-based QRNG using the InP platform. However, the integrated PD is not included in the final measurement due to the high electrical crosstalk between the high-power RF signal driving the diode and the low-power signal reaching the PD. It is worth mentioning that \cite{PRXQuantum.4.010330} with the same method as ours, shows an attractive random number rate of 100\,Gbps relying on efforts made on postprocessing for reducing quantum side-information leakage, to our knowledge, which is the fastest. However, the laser source is bulk instead of integration.

\section{Conclusion}

In this work, we present a miniaturized commercial-ready QRNG entropy which consists of an InP laser diode, a silicon-based homodyne detection module, and electrical amplifier circuits, guaranteeing a shot noise 3\,dB bandwidth of 500\,MHz and a large quantum shot noise clearance of around 20\,dB. 
By sampling with an oscilloscope and extracting random numbers with a computer, a random number generation rate of 6.57\,Gbps can be achieved. Our work addresses the challenge of laser integration and elevated costs associated with QRNG, paving the way for the potential widespread implementation of QRNG in large-scale quantum information networks and secure high-end terminals. Our next phase of work will involve continuous enhancement of the electrical circuit, aligning it with the primary constraints of optical circuits, such as the 40\,GHz 3\,dB bandwidth of the photodetectors employed in our research. By addressing this limitation, we can substantially increase the rate of random number generation. Additionally, we are working on realizing noise signal sampling and post-processing within an electrical integration module, which can be adapted to a broader range of information-secure scenarios.

\section*{Acknowledgements}

\textbf{Funding:} X.H acknowledges the support from Department of Science and Technology of Hubei province (Grant No. 2022BGE009, Grant No. 2024BAB005). Y.Zhang acknowledges National Natural Science Foundation of China (Grant No. 62001044), the Basic Research Program of China (Grant No. JCKY2021210B059), the Equipment Advance Research Field Foundation (Grant No. 315067206), and the Fund of State Key Laboratory of Information Photonics and Optical Communications. B.X acknowledges National Natural Science Foundation of China (Grant No. 62171418) and the Sichuan Science and Technology Program (Grant No. 2023JDRC0017), and Natural Science Foundation of Sichuan Province (Grant No. 2023NSFSC0449).\\
\\
\textbf{Author contributions:} X.H, S.Z. and J.Yan designed and fabricated the device. X.H, Y.B., J.D., and J.Yang characterized the device. X.H and Y.B. wrote the manuscript. Y.Zhang and X.X. supervised this work and provided critical points on the manuscript. Y.Zhu, M.L., D.C., S.Y. and B.X. gave fruitful discussions and revision suggestions. All authors reviewed and approved the final version of the manuscript.\\
\\
\textbf{Conflicts of Interest:} The authors declare no competing interests.

\printbibliography

\end{document}